\documentclass[prd,aps,twocolumn,a4paper,floatfix,showpacs,nofootinbib]{revtex4-1}

\usepackage[utf8]{inputenc}  
\usepackage[T1]{fontenc}
\usepackage{graphicx,psfrag}
\usepackage{mathrsfs}
\usepackage{amsmath,amsfonts,amssymb,pifont,gensymb}
\usepackage{multirow,enumerate}
\usepackage{comment,hyperref}
\usepackage{color}
\usepackage{acronym}
\usepackage{xspace}
\usepackage[normalem]{ulem}
\usepackage{mathtools}
\usepackage{subfigure}
\usepackage{makecell}

\newacro{BH}{black hole}
\newacro{NS}{neutron star}
\newacro{PN}{Post-Newtonian}
\newacro{BBH}{binary black hole}
\newacro{BNS}{binary neutron star}
\newacro{EOB}{effective-one-body}
\newacro{NR}{numerical relativity}
\newacro{GW}{gravitational wave}
\newacro{EOS}{equation-of-state}

\newcommand{\be}{\begin{equation}}
\newcommand{\ee}{\end{equation}}
\newcommand{\bea}{\begin{eqnarray}}
\newcommand{\eea}{\end{eqnarray}}
\newcommand{\bel}{\begin{align}}
\newcommand{\eel}{\end{align}}

\newcommand{\linf}{\texttt{LALInference}}

\newcommand{\imrpnrt}{\texttt{IMRPNRT}}
\newcommand{\imrdnrt}{\texttt{IMRDNRT}}
\newcommand{\imrdnrtqm}{\texttt{IMRDNRT$_{\rm Q}$}}

\def\GMc2{{\rm G M_{\odot} c^{-2}}}

\def\SEOBNRv4T{\texttt{SEOBNRv4T}\xspace}

\def\eqA{\mathrm{Eq}_\mathrm{\chi_{\rm eff}}^{\uparrow \uparrow}}

\def\unA{\mathrm{Un}_\mathrm{\chi_{\rm eff}}^{\uparrow \uparrow}}

\usepackage{color}
\definecolor{cyan}{rgb}{0,0.9,0.9}
\definecolor{orange}{rgb}{0.9,0.5,0}
\definecolor{magenta}{rgb}{1,0,1}
\definecolor{purple}{rgb}{0.8,0.4,0.8}
\definecolor{gray}{rgb}{0.5,0.5,0.5}
\definecolor{mygreen}{rgb}{0.1,0.8,0.1}
\definecolor{darkblue}{rgb}{0.0,0.0,0.6}

\begin{document}

\title{Waveform systematics for binary neutron star gravitational wave signals: 
Effects of spin, precession, and the observation of electromagnetic counterparts} 

\author{Anuradha Samajdar}
\author{Tim \surname{Dietrich}}

\affiliation{Nikhef, Science Park, 1098 XG Amsterdam, The Netherlands}

\date{\today}

\begin{abstract}
Extracting the properties of a binary system emitting gravitational waves 
relies on models describing the last stages of the compact binary coalescence. 
In this article, we study potential biases inherent to current 
tidal waveform approximants for spinning and precessing systems.
We perform a Bayesian study to estimate intrinsic 
parameters of highly spinning binary neutron star systems. 
Our analysis shows that one has to include the quadrupolar 
deformation of the neutron stars due to their rotation once 
dimensionless spins above $\chi \sim 0.20$ are reached, otherwise the extracted 
intrinsic parameters are systematically biased. 
We find that at design sensitivity of Advanced LIGO and Virgo, 
it seems unlikely that for GW170817-like sources a clear imprint of precession 
will be visible in the analysis of the signal employing 
current waveform models. 
However, precession effects might be detectable for unequal mass configurations 
with spins larger than $\chi>0.2$. We finalize our study by investigating possible 
benefits of a combined gravitational wave and electromagnetic detection. 
The presence of electromagnetic counterparts 
help in reducing the dimensionality of 
the parameter space with constraints on the sky location, source distance, and inclination. 
However, we note that although a small improvement in the estimation of the tidal 
deformability parameter is seen in these cases, changes in the intrinsic 
parameters are overall very small.
\end{abstract}

\maketitle

\section{Introduction}
\label{sec:intro}

GW170817 marks a breakthrough in multi-messenger astronomy
and is the first detected gravitational wave (GW) signal 
emerging from the coalescence 
of two neutron stars (NSs). 
This event allowed constraining the expansion rate of the universe~\citep{Abbott:2017xzu}, 
proved that NS mergers are the major cosmic source of r-process elements, e.g. 
\citep{Eichler:1989ve,Rosswog:1998hy,Cowperthwaite:2017dyu,
Smartt:2017fuw,Kasliwal:2017ngb,Kasen:2017sxr,Tanvir:2017pws,
Rosswog:2017sdn,Abbott:2017wuw,Ascenzi:2018mbh},
allowed a precise measurement of the speed difference between GWs 
and light~\citep{GBM:2017lvd}, and
placed constraints on alternative theories of 
gravity~\cite{Ezquiaga:2017ekz,Baker:2017hug,Creminelli:2017sry}. 
In addition, it supported the conjecture that NS 
mergers produce Gamma-Ray-Bursts 
(GRBs)~\citep{Paczynski:1986px,Eichler:1989ve,Monitor:2017mdv}
and enabled the scientific community to investigate 
the supranuclear equation of state (EOS) governing the interior of NSs.
These constraints arise either purely from the analysis of the GW 
signal, e.g., Refs.~\cite{TheLIGOScientific:2017qsa,Dai:2018dca,
De:2018uhw,Abbott:2018wiz,Abbott:2018exr,LIGOScientific:2018mvr},
from a combination of GW and EM information, 
e.g, Refs.~\cite{Radice:2017lry,Bauswein:2017vtn,Coughlin:2018miv,
Radice:2018ozg,Coughlin:2018fis}, 
or from a statistical analysis of a large set of possible 
EOSs, e.g.,~\cite{Annala:2017llu,Most:2018hfd}. 

To extract information from the detected GW signal, 
one needs to cross-correlate the data with 
model waveforms constructed from theoretical predictions. 
Within the framework of Bayesian analysis, this means that a 
multi-dimensional likelihood function has to be 
computed~\cite{Veitch:2014wba}. 
Because of the need to generate 
a large number of GW waveforms for evaluating
the multidimensional likelihood integral numerous times,
the computation of 
each individual waveform has to be sufficiently fast and, on the other hand, 
accurate enough for a precise measurement of the intrinsic source parameters.

Over the last years, there has been progress in modelling 
BNS coalescences by the development of 
improved analytical post-Newtonian (PN) based models, 
e.g.~\cite{Damour:2012yf,Agathos:2015uaa,Jimenez-Forteza:2018buh,
Abdelsalhin:2018reg,Pani:2018inf}, 
state-of-the-art tidal effective-one-body (EOB) waveform approximants, 
e.g., \cite{Bernuzzi:2014owa, Hotokezaka:2016bzh, Hinderer:2016eia, 
Steinhoff:2016rfi, Dietrich:2017feu,Nagar:2018zoe,Nagar:2018plt,Akcay:2018yyh} 
(and their corresponding surrogates~\cite{Lackey:2016krb,Lackey:2018zvw}), 
or closed-form tidal models~\cite{Dietrich:2017aum,Dietrich:2018upm,
Dietrich:2018uni,Kawaguchi:2018gvj}.\\

Bayesian studies characterizing the estimation of the tidal deformability 
and consequently the supranuclear EOS have been presented first
in~\cite{DelPozzo:2013ala} with a PN based waveform model. 
This initial work has been extended and improved in a number of works 
(all based on PN approximants), e.g.,
Refs.~\cite{Agathos:2015uaa,Lackey:2014fwa,Favata:2013rwa,Wade:2014vqa}. 

Relatively recently, Refs.~\cite{Dudi:2018jzn,Messina:2019uby} 
investigated the importance of the inclusion of 
tidal effects for the extraction of the 
NS masses and spins
using non-spinning hybrid tidal EOB -- numerical relativity 
based injections, and the performance of different waveform approximants. 
Ref.~\cite{Abbott:2018wiz} studied tidal EOB waveforms 
(including small effective spins parameters up to $\chi_{\rm eff} \approx 0.02$)
and Ref.~\cite{Samajdar:2018dcx} investigated the imprint of 
the point-particle and tidal description for non-spinning BNS systems on 
parameter estimation (PE) results. 

As per our knowledge, none of the existing works investigated 
the influence of large spins and precession for current 
state-of-the-art waveform approximants, 
or allowed a clear understanding of the effect of a confirmed EM counterpart, 
i.e.,  how different EM observations support the GW data analysis. 
To fill these gaps, we study equal and unequal mass BNS signals for a number 
of spin values and spin orientations, which enable a clear assessment of effects
due to spin and precession. 
In addition, we analyse signals for which we restrict, according to different scenarios,
the source location and/or inclination of the binary. 

The article is structured as follows. 
In Sec.~\ref{sec:methods} we discuss briefly the numerical methods and 
employed waveform approximants. 
In Sec.~\ref{sec:spins} we investigate the effect of aligned spin and precession 
on the parameter estimation analyses. Sec.~\ref{sec:EM} gives a short 
overview about the effects of possible observed EM counterparts and 
their imprint on the GW data analysis. 
We summarize and conclude in Sec.~\ref{sec:conclusion}.

\section{Methods}
\label{sec:methods}

\subsection{Bayesian inference}
In this article, we use a Bayesian approach for parameter estimation based on the 
\linf~module~\cite{Veitch:2014wba} available in the \texttt{LALSuite} package. 
In particular, we employ the Markov Chain Monte Carlo (MCMC) algorithm 
\texttt{lalinference\_mcmc}~\cite{Rodriguez:2013oaa}. 
Information about the parameters are encoded in the 
posterior probability distribution function 
\begin{equation}
 p(\vec{\theta}|\mathcal{H}_s,d) = \frac{p(\vec{\theta}|\mathcal{H}_s)p(d|\vec{\theta},\mathcal{H}_s)}{p(d|\mathcal{H}_s)},
 \label{eqn:Bayes}
\end{equation}
where $\vec{\theta}$ represents the parameter set and $\mathcal{H}_s$ the hypothesis that a 
GW signal depending on the parameters $\vec{\theta}$ is present in the data $d$.
In addition to the parameter set common to a BBH signal 
$\{ m_1, m_2, \chi_1, \chi_2, \theta, \phi, \iota, \psi, D_L, t_c, \varphi_c \}$, 
the two tidal deformability parameters $\Lambda_A$ and $\Lambda_B$ are present 
for a BNS system, see e.g.~\cite{Samajdar:2018dcx} for further details.  
The best measured quantity describing tidal effects is in fact a mass-weighted combination of the 
individual tidal deformabilities, e.g.~\cite{Flanagan:2007ix}, 
\begin{equation}
\tilde{\Lambda}=\frac{16}{13} \sum_{i=1,2} \Lambda_i \frac{m_i^4}{M^4}\left( 12-11\frac{m_i}{M} \right). 
\end{equation}

The likelihood of obtaining a signal $h(t)$ in data stream $d(t)$, 
which also includes the noise $n(t)$, is proportional to 
\begin{equation}
p(d|\vec{\theta},\mathcal{H}_s) \propto \exp{\left [-\frac{1}{2}(d-h|d-h)\right ]}. 
\end{equation}

\subsection{Waveform models}
\label{subsec:wfrms}
A frequency-domain gravitational waveform is given by
\begin{equation}
 \tilde{h}(f) = \tilde{A}(f) e^{-i \Psi(f)},
\end{equation}
where the phase $\Psi(f)$ can be approximated as 
a sum of the non-spinning point-particle (PP) contribution, 
a spin-orbit (SO) contribution, a spin-spin (SS) 
contribution, and tidal contribution (Tides): 
\begin{equation}
 \Psi(f) = \Psi_{\rm PP}(f) + \Psi_{\rm SO}(f) + \Psi_{\rm SS}(f) +  \Psi_{\rm Tides}(f). 
 \label{eq:totPhase}
\end{equation}
The PE analyses of the BNS signal GW170817 
in~\cite{Abbott:2018wiz,Abbott:2018exr,LIGOScientific:2018mvr} have also been 
done with BNS waveform models where the tidal phasing given by the 
\texttt{NRTidal} framework~\cite{Dietrich:2017aum} 
is added to the BBH inspiral-merger-ringdown 
waveform models \texttt{IMRPhenomD}~\cite{Khan:2015jqa},
\texttt{IMRPhenomPv2}~\cite{Schmidt:2014iyl}, 
and \texttt{SEOBNRv4\_ROM}~\cite{Bohe:2016gbl,Puerrer:2014fza}. 
Details of construction of these BNS waveform models 
from their BBH counterparts are presented 
in~\cite{Dietrich:2018uni}.

In this article, we restrict the analysis to three waveform models:
\begin{enumerate}[(i)]
 \item The spin-aligned \texttt{IMRPhenomD\_NRTidal} (henceforth \imrdnrt) 
       model as discussed in~\cite{Dietrich:2018uni}. This model does not include 
       EOS-dependent quadrupole-monopole terms, i.e., all spin-spin contributions
       are treated similar to that of a binary black hole system.
 \item The precessing \texttt{IMRPhenomPv2\_NRTidal} (henceforth \imrpnrt) 
       model as discussed in~\cite{Dietrich:2018uni}. This model includes the 
       EOS-dependent quadrupole-monopole term up to the 3PN order. 
 \item The spin-aligned \texttt{IMRPhenomD\_NRTidal} for which we add 
       the EOS dependence to the spin-spin interactions (henceforth \imrdnrtqm) 
       up to the 3PN order. 
\end{enumerate} 

The main spin contribution is characterized by the aligned-spin effective 
parameter
\begin{equation}
 \chi_{\rm eff} = \frac{m_1 \chi_{1_\parallel} + m_2 \chi_{2_\parallel}}{m_1+m_2},
\end{equation}
where $\chi_{i_\parallel} = \frac{\vec{S}}{m_i^2}\cdot \hat{L}$ is the dimensionless
spin parameter aligned with the direction of the orbital angular
momentum $\hat{L}$. Precession effects in \imrpnrt\ are in addition 
characterized by the 
single-spin parameter~\cite{Schmidt:2012rh} 
\begin{equation}
 \chi_{\rm p} = {\rm max} \left( \chi_{1_\perp}, 
 \frac{3+4q}{4+3q} q \ \chi_{2_\perp} \right),
\end{equation}
where $\chi_{1_\perp}$ and $\chi_{2_\perp}$ denote the spin components 
perpendicular to the orbital angular momentum and $q$ is the mass-ratio 
defined below in Sec.~\ref{subsec:inj_study}. 
We have investigated effects of varying $\Psi_{\rm PP}(f)$
and $\Psi_{\rm Tides}(f)$ in Ref.~\cite{Samajdar:2018dcx}. 
Here, we focus on the effects of 
$\Psi_{\rm SO}(f)$, $\Psi_{\rm SS}(f)$, precession, 
as well as possible improvements in the 
PE analyses due to a more constrained parameter space 
from accompanying EM counterparts.

\subsection{Injection study} 
\label{subsec:inj_study}

\begin{table}
\renewcommand{\arraystretch}{1.20}
\begin{tabular}{c |c| c| c | c | c|c |c}
 Name & $m_1$ & $m_2$ & $\mathrm{M}_\mathrm{total}$ & $\mathrm{q}$ & $\Lambda_1$ & $\Lambda_2$ & $\tilde{\Lambda}$ \\
\hline
 Eq & 1.375  & 1.375 & 2.75 & 1.00 & 292 & 292 & 292 \\
 Un & 1.68 & 1.13 & 2.81 & 0.67 & 77 & 973 & 303  \\       
\end{tabular}
\caption{Overview of the intrinsic parameters of the injected sources, 
the individual spins are varied in addition. 
The columns refer to the individual masses, the total mass, the
mass ratio, the individual tidal deformabilities, 
and the effective tidal deformability $\tilde\Lambda$.}
\label{tab:sources_aligned_spin}
\end{table}

As an EOS which is in agreement with current constraints~\cite{Abbott:2018wiz}, 
we choose the component masses and tidal deformabilities following the APR4 
EOS~\cite{Read:2008iy}.
We place the sources at a luminosity distance of $D_L = 50 \ \mathrm{Mpc}$.
The intrinsic parameters of the two employed BNS sources are 
summarized in Tab.~\ref{tab:sources_aligned_spin}. We choose values of the 
dimensionless spin parameter varying from $0.05$ to $0.50$.
We simulate spinning BNS sources and investigate biases in estimation of 
the binary parameters in presence of three different spin configurations
\begin{enumerate}[(i)]
 \item spins aligned with the direction of the orbital angular 
       momentum (aligned spins) -- denoted as ($\uparrow \uparrow$);
 \item spins lying in the orbital plane (in-plane spins) 
       -- denoted as $(\leftarrow \ \rightarrow)$;
 \item spins oriented at an angle of $45^\circ$ with the direction of 
       the orbital angular momentum (misaligned spins) -- denoted as 
       ($\nwarrow \ \nearrow$).
\end{enumerate}
Our simulations are done using the waveform model 
\imrpnrt\ and are analyzed with \imrdnrt, \imrpnrt, and \imrdnrtqm.
All simulations are done in simulated, Gaussian noise using noise power spectral 
density of the design sensitivity of  
Advanced LIGO and Virgo's detector network~\cite{lvc_psds}.
We note that our simulations are quite loud with matched filter
signal to noise ratios of 80-90, so our results are dominated mainly by systematics rather 
than statistical uncertainties.

Our priors are motivated by the study of GW170817, Ref.~\cite{Abbott:2018wiz}.
Consequently, the recovery is done with a uniform prior on 
the component tidal deformabilities $\Lambda_1$ and $\Lambda_2$ between 0 and
5000.  
Priors on dimensionless spin magnitudes are distributed uniformly between 0 and 0.7. 
The chirp mass $\mathcal{M}=(m_1 m_2)^{3/5}/M^{1/5}$, 
is sampled uniformly between 1.184 and 2.168 with 
mass-ratio $q=m_2/m_1$ restricted between 0.125 and 1. 
Priors on the other parameters are given explicitly in the following sections 
according to the individual analysis performed.

\begin{figure*}[t]
\includegraphics[keepaspectratio,width=0.98\textwidth]{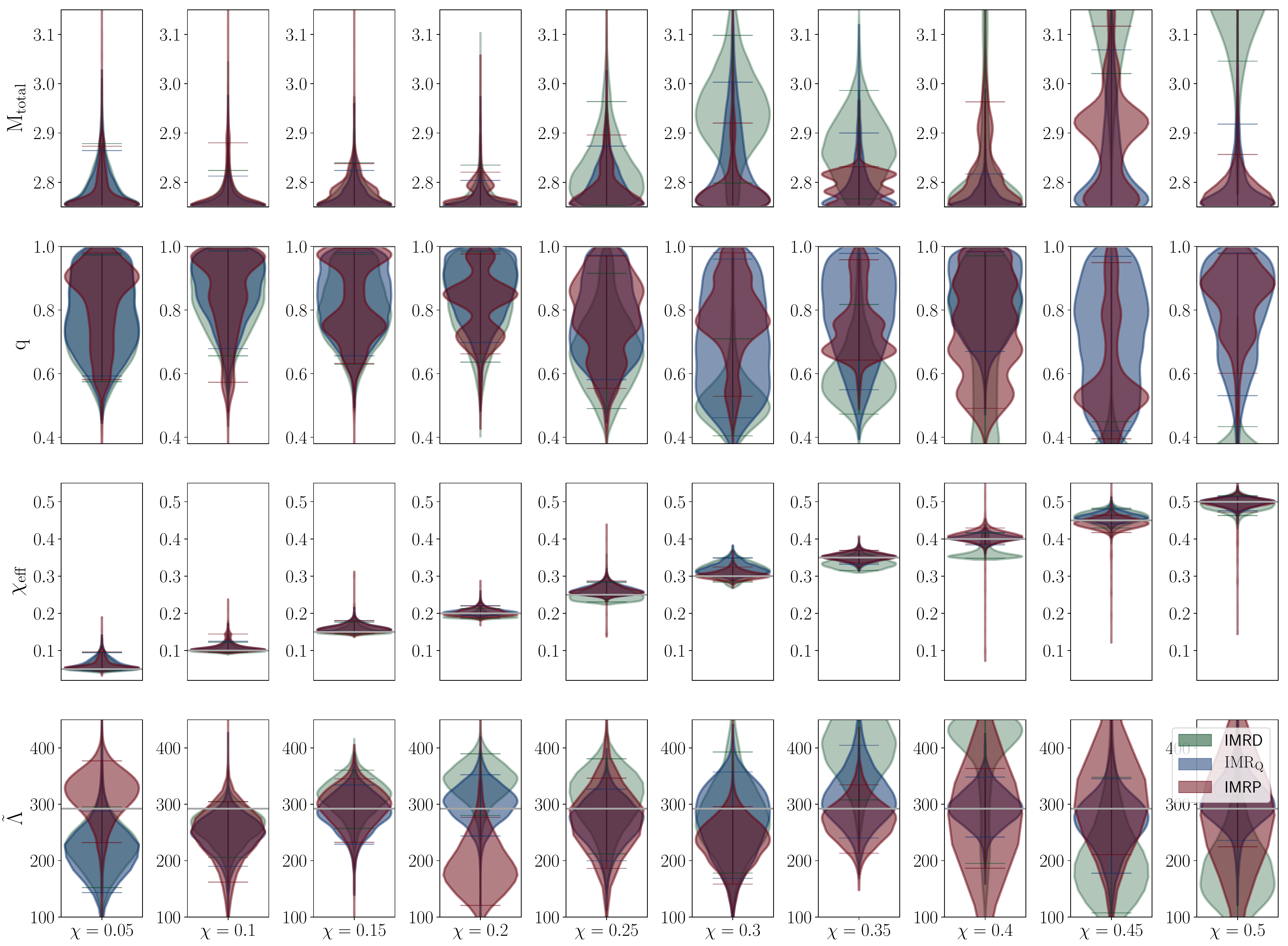}
  \caption{Equal mass configuration $\eqA$. 
  The extracted posteriors are marked green for the \imrdnrt\ approximant, red for \imrpnrt, 
  and blue for \imrdnrtqm. 
  Injected value are marked with a horizontal gray line and  
  $90\%$ confidence intervals are shown with green, red, and blue horizontal lines 
  for the individual approximants.}
  \label{fig:violin_no_em_equal}
\end{figure*}

\begin{figure*}[t]
\includegraphics[keepaspectratio,width=0.98\textwidth]{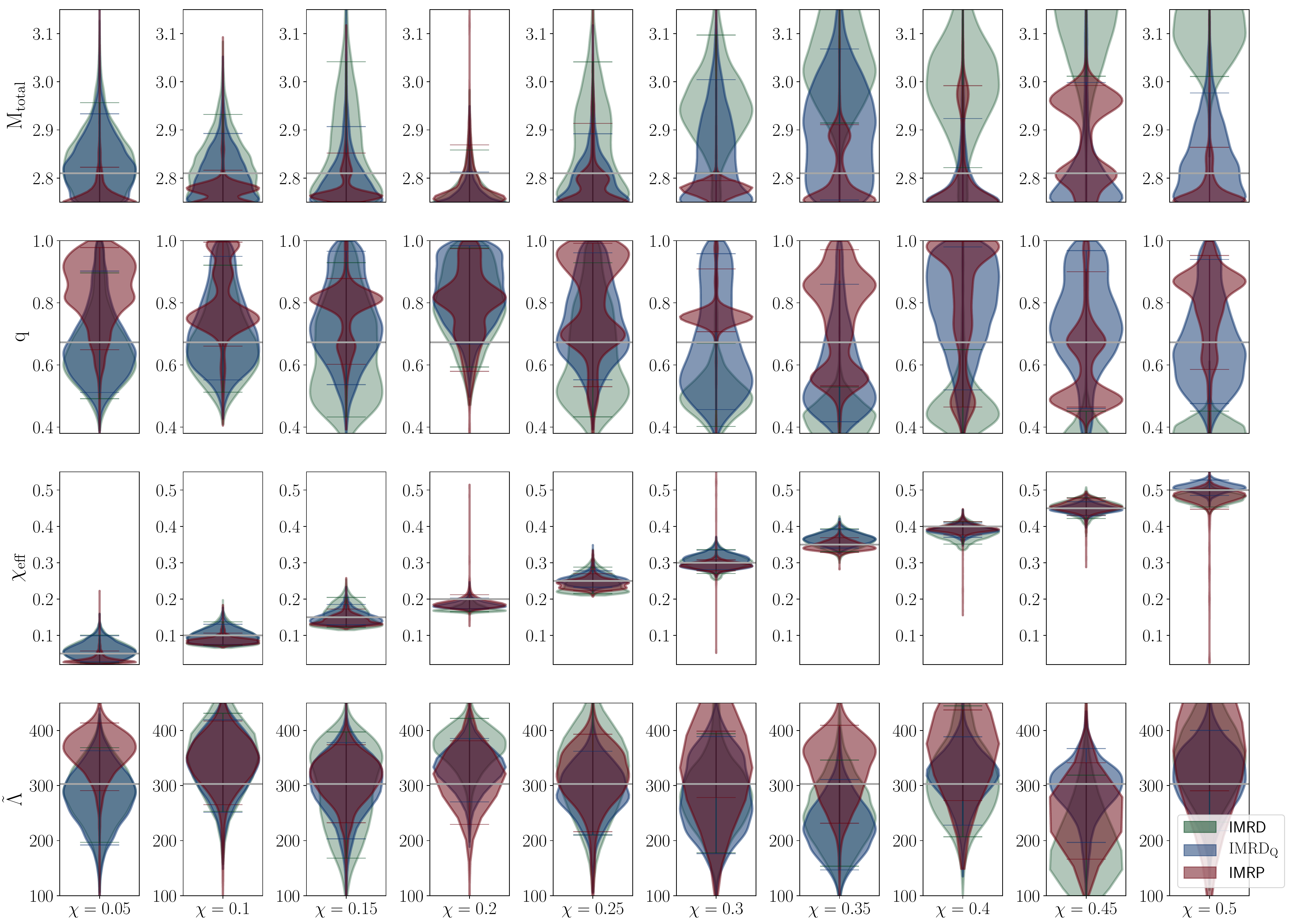} 
  \caption{As Fig.~\ref{fig:violin_no_em_equal} but for the unequal-mass configurations $\unA$.}
  \label{fig:violin_no_em_unequal}
\end{figure*}

\begin{figure*}[t]
\includegraphics[width=0.46\textwidth]{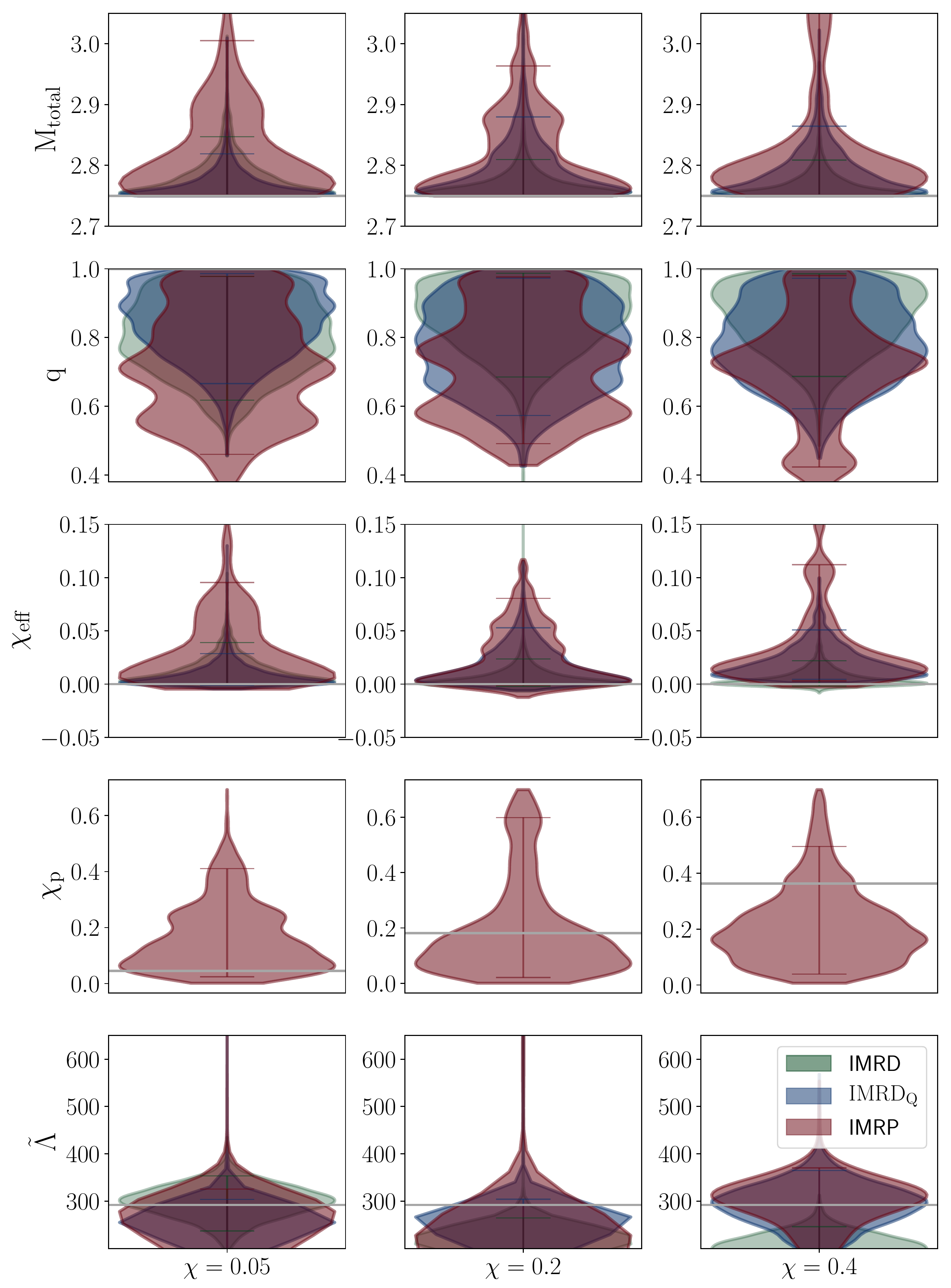} \hfill
\includegraphics[width=0.46\textwidth]{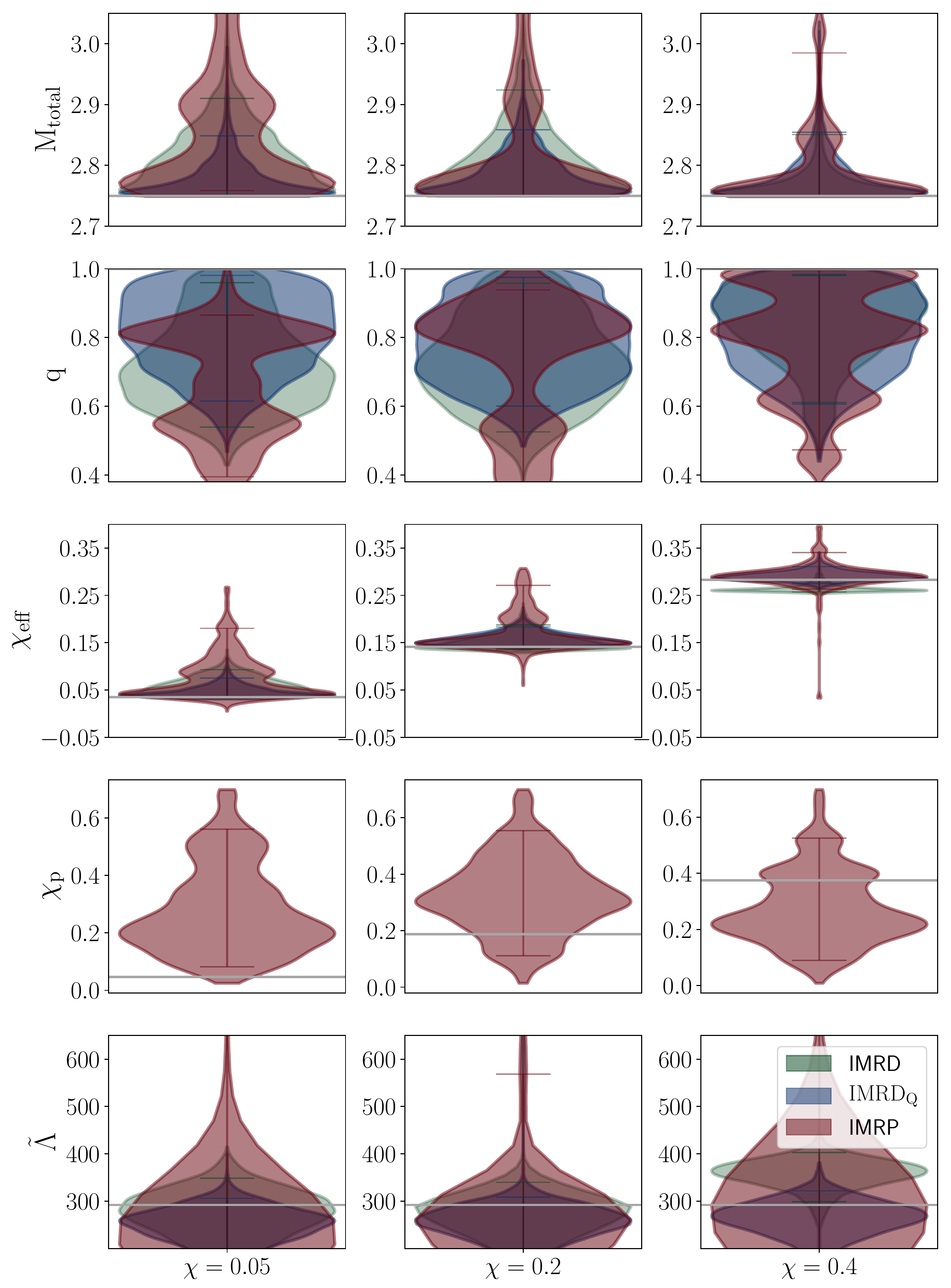}
\caption{Equal mass system $\mathrm{Eq}_\mathrm{x}$ with in in-plane spin $({\leftarrow \rightarrow})$ (left)
         and with mis-aligned spin $({\nwarrow \nearrow})$ (right). 
         Injected values are marked with a horizontal gray line and $90\%$ confidence intervals
         are shown with red/green horizontal lines. }
  \label{fig:res_spin_precess_eq_mass}
\end{figure*}

\begin{figure*}[t]
\includegraphics[width=0.46\textwidth]{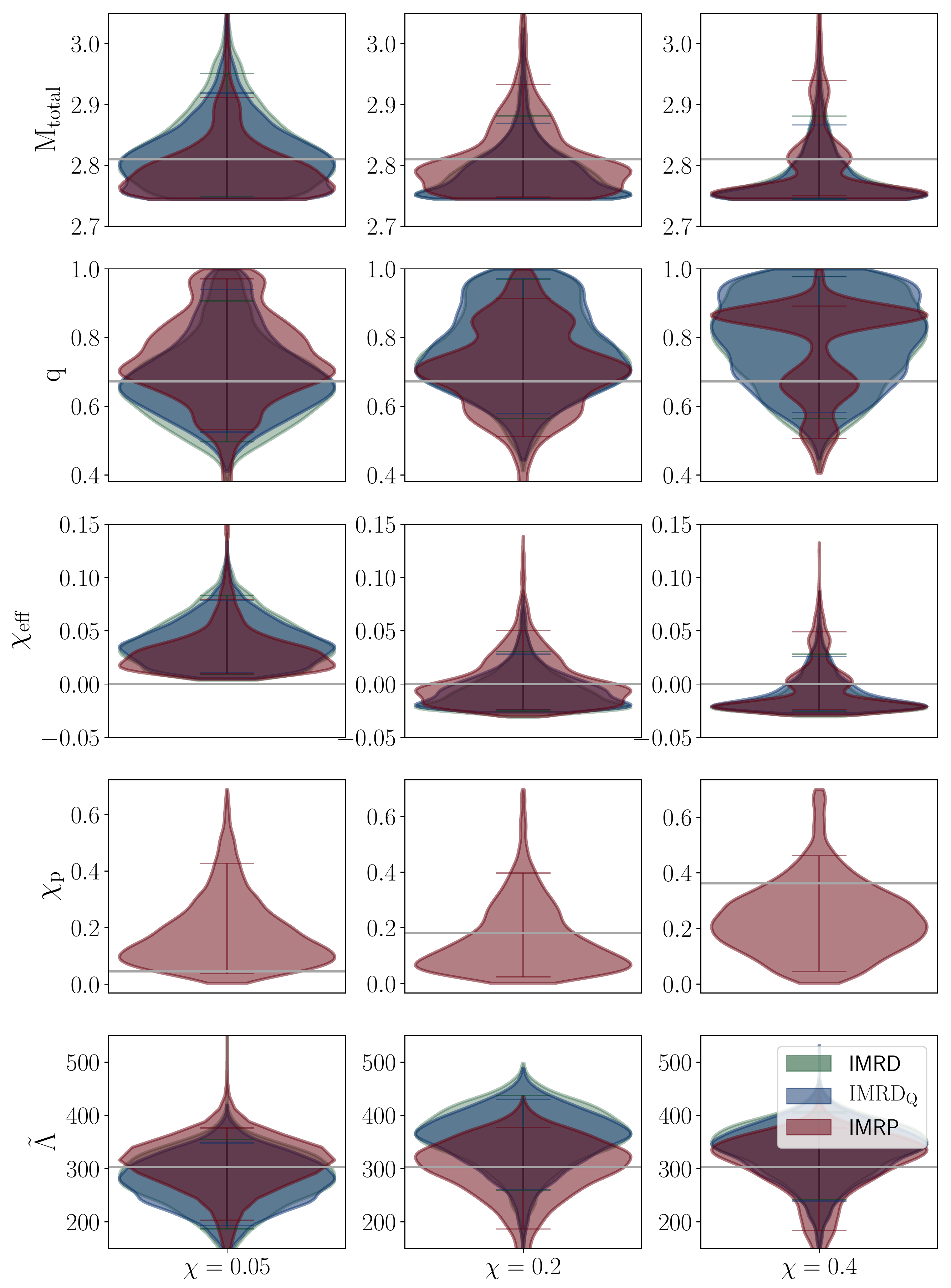} \hfill
\includegraphics[width=0.46\textwidth]{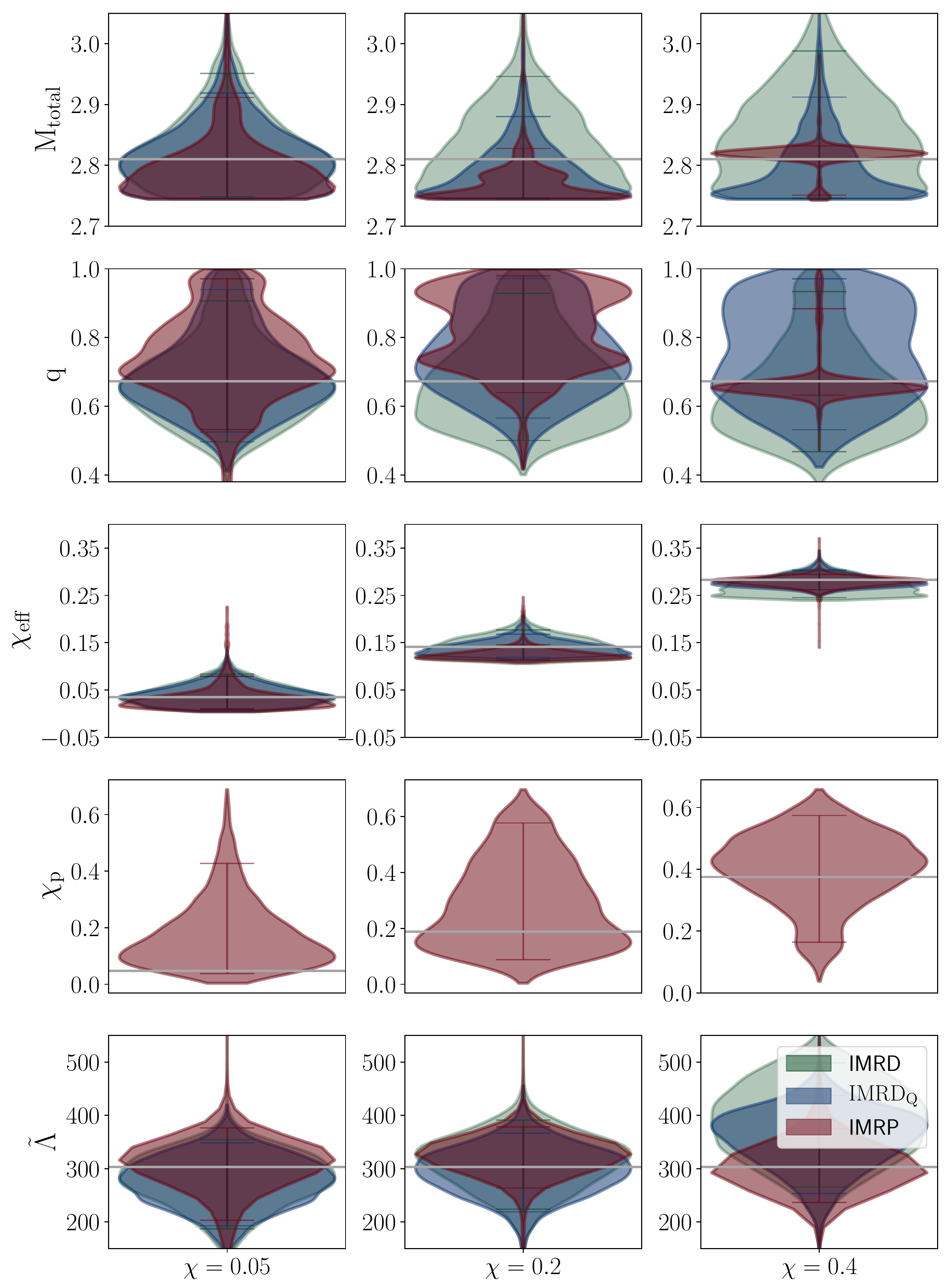}
\caption{Unequal mass system $\mathrm{Un}_\mathrm{x}$ with in in-plane spin ${(\leftarrow \rightarrow})$ (left)
         and with mis-aligned spin $({\nwarrow \nearrow})$ (right). 
         Injected values are marked with a horizontal gray line and $90\%$ confidence intervals
         are shown with red/green horizontal lines. }
  \label{fig:res_spin_precess_uneq_mass}
\end{figure*}

\section{Extracting intrinsic parameters in presence of spins}
\label{sec:spins}

In this section, we simulate equal-mass and unequal-mass 
sources by varying their dimensionless spin magnitudes. 
Apart from their spins, the intrinsic parameters of each of the 
equal and unequal mass source are kept fixed and 
are listed in Tab.~\ref{tab:sources_aligned_spin}.

In addition to the priors mentioned above (Sec.~\ref{subsec:inj_study}), 
we sample the distance uniformly in a co-moving volume up to $100\rm Mpc$.
The priors on the sky position as well as the inclination of the binary are 
uniformly distributed on the sphere. 

\subsection{Effects of aligned spins}
\label{subsec:aligned_spin}

For each of the sources in Tab.~\ref{tab:sources_aligned_spin}, 
we vary the dimensionless spin magnitudes on each component mass
from $0.05$ to $0.5$ in steps of $0.05$. 
Henceforth, we refer to an aligned-spin equal-mass system 
with effective spin parameter $\chi_{\rm eff}$ 
as $\eqA$. Similarly, the unequal-mass aligned-spinning 
systems are denoted by $\unA$.\\

\textbf{Equal-mass binaries:}

In Fig.~\ref{fig:violin_no_em_equal}, we show the posterior 
probability distribution functions (PDFs) of the intrinsic 
parameters for each of the injected spin values
for the equal-mass setup.
Considering the extraction of intrinsic binary properties from the equal-mass binary, 
we find generally that with \imrpnrt\ or \imrdnrtqm\ the binary parameters are better 
recovered than for the corresponding \imrdnrt\ setup. 
This points to the importance of the spin-spin interactions,
in particular for spins $\chi>0.2$. 

Independent of the model, for an increasing spin magnitude,
the extraction of the total mass, mass ratio, 
and tidal deformability becomes less restrictive. 
This is most visible for \imrpnrt\ when considering the tidal deformability parameter 
$\tilde\Lambda$. Although the injected value (horizontal gray lines for each panel 
in Fig.~\ref{fig:violin_no_em_equal}) lies almost always within 
the 90\% credible interval (given by the colored horizontal lines; the colors 
corresponding to each waveform approximant), 
the posterior distribution becomes increasingly broad for larger spins.
In comparison, the \imrdnrtqm\ model restricts the tidal deformability better 
than \imrpnrt. A similar effect was discussed in Ref.~\cite{Samajdar:2018dcx} for
non-spinning injections, where it was pointed that the reduction of the parameter 
space helps to improve the extraction of individual parameters. 
Trivially, this observation can be generalized to aligned-spin configurations. 

With all other parameters being kept fixed, 
a larger value of $\chi_{\rm eff}$ causes
a longer inspiral phase~\cite{Campanelli:2006uy,Bernuzzi:2013rza,
Dietrich:2016lyp}. 
It has been shown~\cite{Ohme2012} 
that effects of larger spin may be compensated by changing 
values of the mass ratio~\cite{Baird:2012cu}
or even total mass. 
We do see a similar trend in the recovered mass ratios as they become smaller 
for larger spin magnitudes. 
For these cases, the estimated total masses are higher, 
which leads to a recovered chirp mass closer to the injected value.

Among the best recovered parameters for all models is 
the effective aligned-spin parameter $\chi_{\rm eff}$. 
This can be understood by the fact that the effective spin is 
mainly determined from 
the long early-inspiral containing several thousand GW cycles, 
since the leading order spin-orbit contribution enters already 
at the 1.5PN order~\footnote{We note that although the mass ratio 
enters at even earlier PN order, it is measured better at higher frequencies, 
which follows e.g.~from Fig.~2 of~\cite{Harry:2018hke} and is in agreement with our results}. 
Only the \imrdnrt\ approximant shows 
noticeable differences with respect to the injected values, 
which again emphasizes the importance of incorporation of the 
spin-induced quadrupole-monopole term. 

It is encouraging that standard waveform approximants are able 
to determine spin values of $\sim 0.05$ once the design sensitivity of advanced GW detectors is reached. 
This shows that there is the chance of 
detecting millisecond pulsars via GW astronomy. 

In general, it seems possible to reliably estimate intrinsic parameters 
up to an aligned spin of $\chi_{\rm eff} \sim 0.4$ once the 
spin-spin interactions include EOS dependence as in \imrdnrtqm\ and \imrpnrt. 
We note that this value is well above the largest dimensionless spin 
observed in a BNS system to date ($\chi \sim 0.05$ for PSR J1946+2052~\cite{Stovall:2018ouw}), 
and therefore is also above the low-spin prior of $\chi<0.05$ 
employed in some of the LVC analyses, 
e.g.~\cite{Abbott:2018wiz,Abbott:2018exr,LIGOScientific:2018mvr}. 
It is also larger than the dimensionless spin of PSR J1807$-$2500B 
which is, with a rotation frequency of $239$Hz \cite{Lorimer:2008se,Lattimer:2012nd}, 
currently the fastest spinning NS in a binary system. \\

\textbf{Unequal-mass binaries:}

In Fig.~\ref{fig:violin_no_em_unequal}, we show the posterior 
probability distribution functions (PDFs) of the intrinsic 
parameters for each of the injected spin values
for the unequal-mass setup.
The results for the unequal-mass configurations are in line with our equal-mass studies. 
We find that \imrpnrt\ and \imrdnrtqm\ approximants perform best and that above a 
spin value of $\chi \sim 0.25$ the quadrupolar deformation of the NSs due to its 
own spin becomes important. 
Thus, systematic biases are introduced in the \imrdnrt\ model in which the 
EOS-dependent quadrupolar deformation is not included. 
These biases result in a larger estimated total mass and 
smaller mass ratio (as for the equal-mass case).

Considering the effective spin $\chi_{\rm eff}$, we find that as for the equal mass 
case the injected value (denoted by horizontal gray lines in each panel in Fig.~\ref{fig:violin_no_em_unequal}) 
is recovered robustly and always remains within the 90\% credible interval 
(colored horizontal lines, with the colors 
corresponding to each waveform approximant). 
The injected value of $\tilde{\Lambda}$ is always contained
within the posterior distribution of the $\tilde{\Lambda}$ parameter for all waveform approximants. 
As for the equal-mass case, 
the posterior distribution functions with \imrpnrt\ recovery broaden with gradual increase in spins.
However, the same does not hold true for \imrdnrtqm\ recovery.

\subsection{Effects of precessing spins}

We now consider systems with precession, i.e., 
setups for which a non-vanishing spin component inside the orbital plane exists. 
To reduce computational costs, we study only setups for the spin values 
$\chi=0.05,0.2,0.4$ and discard other spin values studied for the aligned cases. 

\textbf{Equal-mass binaries:}
The results from the precessing spin equal-mass simulations are shown in 
Fig.~\ref{fig:res_spin_precess_eq_mass}. We show in the left panel results 
for the in-plane configuration and in the right panel results for the misaligned setup.

For the in-plane configuration, the simulated values of total mass $M$ and mass ratio $q$ lie within the $90\%$ upper bound values 
of their posterior distributions. 
The effective spin $\chi_{\rm eff}$ 
is also recovered within the $90\%$ confidence region for in-plane systems although the confidence interval 
is quite broad to be able to make conclusive statements. 
For all values of spin magnitudes, recovery with the \imrpnrt\ model includes the injected value of 
the precessing spin parameter $\chi_{\rm P}$ within the $90\%$ confidence interval 
of the posterior distribution. However, the confidence intervals 
are generally too large to claim confidently a measurement of precession effects.  
When the injected spin magnitude is large ($\chi \sim 0.4$), 
we note that \imrdnrt\ performs 
poorly to recover the extracted tidal deformability whereas 
\imrdnrtqm\ and \imrpnrt\ perform similarly, in spite of \imrdnrtqm\ being an aligned spinning model. 

In an improved model of the \texttt{NRTidal} framework~\cite{Dietrichinprep}, 
higher order spin-squared and spin-cubed terms are added, along with their respective 
spin-induced moments. This might improve the estimation of the tidal deformability 
in the case of large spins.

For the misaligned configurations (right panel of Fig.~\ref{fig:res_spin_precess_eq_mass}), 
the total mass $M$ is recovered well with \imrdnrt, 
but for large spins we find similar biases as for the aligned configuration 
for \imrdnrt. The $\chi_{\rm eff}$ parameter however, seems to be quite robustly measured.
We find the tidal deformability to be loosely constrained with \imrpnrt\ 
for this spin orientation.

Overall, our observations might hint towards the fact that a further improvement 
of the precession dynamics is needed. A possible extension of the 
\imrpnrt\ model with a two-spin precession description as in~\cite{Khan:2018fmp} 
might help to further improve our capability to extract precession effects from future events. 
On the other hand, while computationally more expensive, one could also extend the 
precessing EOB model~\cite{Pan:2013rra} with the \texttt{NRTidal} framework to allow a better modelling of 
the inspiral of precessing BNSs. 

\textbf{Unequal-mass binaries:}
The results from the simulations of the unequal-mass setup for in-plane (left panel)
and misaligned spin (right panel) are shown in Fig.~\ref{fig:res_spin_precess_uneq_mass}.
All waveform models include the injected parameters within the $90\%$ 
confidence intervals of the posterior distributions, although it appears again that $\chi_{\rm eff}$ is slightly 
overestimated for the $\chi=0.05$ configuration. 
While most observations made for the equal-mass sources hold also for the unequal-mass 
sources, we find that for very large misaligned spins ($\chi =0.4$  and the $({\nwarrow \nearrow})$ case) 
the \imrpnrt\ posterior of $\chi_p$ excludes non-precessing setups. 
This shows that, as expected, precession effects are easier to detect 
for unequal mass setups than for equal mass systems. 

\section{Incorporating information from electromagnetic counterparts}
\label{sec:EM}

In addition to the study of spin effects, presented in the 
previous section, we want to understand the possible interplay 
between GW astronomy and EM observations. 
For this purpose, we will employ the configurations 
$\mathrm{Un}_\mathrm{\chi}^{(\uparrow \uparrow)}$, 
$\mathrm{Un}_\mathrm{\chi}^{(\leftarrow \rightarrow)}$,
$\mathrm{Un}_\mathrm{\chi}^{(\nwarrow \nearrow)}$
with $\chi=0.2$ and $0.4$, i.e., 
a total of six different physical setups.
To save further computational costs, 
we restrict our analysis to the \imrpnrt\ model since it is the 
only precessing model and analyse for the observational scenarios II and III 
(kilonova and GRB respectively, cf. Tab.~\ref{tab:EM_priors}) only the $\chi=0.2$ case and with an 
aligned-spinning configuration.\\

GW170817~\cite{GBM:2017lvd} has shown the huge variety of EM signals 
which can be detected in coincidence or as a follow-up of a potential GW trigger. 
We consider here four different scenarios:\\

\noindent \textbf{I. the absence of an EM counterpart:} For such scenarios one keeps 
 the GW standard priors on the distance and sky location.\\
 
 \noindent  \textbf{II. a kilonova detection:} In cases where a kilonova gets detected, the 
 good angular resolution of optical or near-optical telescopes fixes the sky location 
 and provides, due to the redshift measurement of the host galaxy, a constraint on the 
 source distance. \\
 
 \noindent  \textbf{III. a GRB detection:} The observation of a GRB generally provides broad information 
 about the sky location. Furthermore, since GRBs are beamed within a small angle, they also provide additional 
 estimates of the inclination of the binary with respect to the line of sight, 
 e.g.,~\cite{Eichler:1989ve,Paczynski:1991aq,Narayan:1992iy,Nathanail:2018jpu,vanEerten:2018vgj,Wu:2018bxg}.\\
 
\noindent   \textbf{IV. kilonova and GRB detection:} As for GW170817 it can be expected that for a number of BNS 
 mergers, one detects both a kilonova and a GRB which combines the constraints of (ii) and (iii). \\

We provide for all four cases the priors for the distance, sky localization ($\theta,\phi$), 
and inclination of the system $\iota$ in Tab.~\ref{tab:EM_priors}.

\begin{table}
\renewcommand{\arraystretch}{1.20}
\begin{tabular}{c | cccc}
 \multicolumn{1}{c|}{Counterpart} & \multicolumn{4}{c}{Priors} \\ \hline
 & $D_L \ [\mathrm{Mpc}]$ & $\theta \ [\mathrm{deg.}]$ & $\phi \ [\mathrm{deg.}]$ & $\iota \ [\mathrm{deg.}]$ \\
\hline
 Injected values & 50 & 60 & 60 & 25 \\
 No EM counterpart & [1,100] & [0,180] & [-90,90] & [$0$, 180]  \\
 Kilonova & [45,55] & fixed & fixed & [$0$, 180] \\
 GRB & [1,100] & [35,85] & [35,85]  & [0, 50] \\
 Kilonova + GRB & [45,55] & fixed & fixed & [0, 50]
\end{tabular}
\caption{Overview about the prior choices for different observational scenarios considering 
absence of an EM counterpart, a detection of a kilonova, a detection of a GRB, and the combined 
detection of a kilonova and a GRB. }
\label{tab:EM_priors}
\end{table}

For the interpretation of the imprint of the combined EM and GW analyses, we compute 
the systematic bias with the standard accuracy statistic (stacc) defined as
\begin{equation}
 S = \sqrt{\frac{1}{N} \sum_{i=1}^N (x_i - x_{\mathrm{inj}})^2},
 \label{eq:stacc}
\end{equation}
where $N$ is the number of samples, $x_i$ is the $i^{th}$ posterior sample and $x_{\mathrm{inj}}$ 
is the injected value of the parameter $x$.

We find that although the covered parameter space is reduced due to the additional EM information, 
there are only minor changes in the recovered estimates and the incorporation of the source distance, location, and inclination 
does not noticeably improve the GW data analysis. 
In fact, there is no obvious systematic improvement in measurement of 
$\mathrm{M_{total}}$, $\mathrm{q}$ or $\mathrm{\chi_{eff}}$. 
The only relevant improvement is a better estimate of the 
tidal deformability. 
Table~\ref{tab:stacc_results} shows that for the high spinning configurations 
if some spin contribution is aligned to the orbital angular momentum, 
i.e., setups $(\uparrow \uparrow)$ or $({\nwarrow \nearrow})$, 
the stacc value can be reduced by an order of a few if 
additional EM information are incorporated. 

\begin{table}
\begin{tabular}{l| c| c | c | c}
\hline
  EM scenario & \multicolumn{2}{c|}{Spin configuration} 
  & Stacc  \\ \hline
     & Orientation & Spin magnitude &  $\tilde{\Lambda}$ \\
\hline
 no EM              & $(\uparrow \uparrow)$ & 0.20 & 126.74 \\ 
 no EM              & $(\uparrow \uparrow)$ & 0.40 & 351.70 \\ 
 kilonova           & $(\uparrow \uparrow)$ & 0.20 & 99.40  \\ 
 GRB                & $(\uparrow \uparrow)$ & 0.20 & 56.22  \\ 
 kilonova \& GRB    & $(\uparrow \uparrow)$ & 0.20 & 45.71  \\ 
 kilonova \& GRB    & $(\uparrow \uparrow)$ & 0.40 & 48.13   \\ 
 \hline
 no EM              & $({\leftarrow \rightarrow})$ & 0.20  & 59.27 \\
 no EM              & $({\leftarrow \rightarrow})$ & 0.40  & 59.81 \\ 
 kilonova \& GRB    & $({\leftarrow \rightarrow})$ & 0.20  & 109.72  \\             
 kilonova \& GRB    & $({\leftarrow \rightarrow})$ & 0.40  & 69.15  \\             
 \hline 
 no EM              & $({\nwarrow \nearrow})$ & 0.20  & 74.06 \\
 no EM              & $({\nwarrow \nearrow})$ & 0.40  & 110.06 \\ 
 kilonova \& GRB    & $({\nwarrow \nearrow})$ & 0.20  & 50.81  \\             
 kilonova \& GRB    & $({\nwarrow \nearrow})$ & 0.40  & 44.37  \\     
 \end{tabular}
\caption{Stacc estimtes for different scenarios of combined EM and GW analyses using the \imrpnrt\ model and the unequal-mass source.}
\label{tab:stacc_results}
\end{table}

\section{Summary}
\label{sec:conclusion}

In this work, we have presented a first study to 
classify potential biases in the extraction of binary parameters 
for spinning and precessing GW signals and how detected EM counterparts can support 
the GW parameter estimation.
Our main findings are summarized as follows: 
\begin{enumerate}[(i)]
 \item Considering aligned spinning configurations, the effective spin $\chi_{\rm eff}$ is well constrained 
       for all employed waveform approximants. 
       Thus, we might be able to extract information about the spins of the stars 
       from future BNS events.
 \item Once the star's spins exceed values above $\chi \sim 0.2$, 
       GW models need to contain the EOS dependent quadrupole-monopole contribution in the 
       GW phase description, otherwise results are systematically biased. 
 \item It is unlikely that with the current waveform approximants, precession effects 
       will be detected unless the configuration consists of NSs with different masses 
       and very high spin values. 
 \item Incorporating additional EM information does not lead to a 
       noticeable improvement of the results, only the tidal deformability 
       changes for some of our investigated configurations. 
\end{enumerate}

While this work has only been the first step towards a better 
understanding of waveform model systematics for spinning, tidal 
models, it already hints towards possible improvements for future developments. 
Most notably, we propose the extension of the fully precessing, phenomenological 
BBH model presented in~\cite{Khan:2018fmp} to potentially improve our ability 
to find clear imprints of precession effects from future detections. 
Furthermore, the noticeable importance of the spin-spin interactions 
suggest that a more detailed modelling of these effects as presented 
in~\cite{Nagar:2018plt} for the EOB model or in~\cite{Dietrichinprep}
for the \texttt{NRTidal} approximation, should be the focus of 
future BNS model developments.

\begin{acknowledgments}
  We thank Katerina Chatziioannou, Sebastian Khan, and Chris Van Den Broeck
  for helpful discussions and continuous support.
  
  AS and TD are supported by the research programme 
  of the Netherlands Organisation for Scientific Research (NWO).
  TD acknowledges support by the European Union’s Horizon 
  2020 research and innovation program under grant
  agreement No 749145, BNSmergers. We are grateful for the 
  computing resources provided by
  the LIGO-Caltech Computing Cluster where our simulations 
  were carried out. We are grateful for computational resources provided by 
  Cardiff University, and funded by an STFC grant supporting UK Involvement in the Operation of Advanced
LIGO.
\end{acknowledgments}

\bibliography{spin_pn_nr_tides}

\end{document}